\documentclass[aps,twocolumn,superscriptaddress]{revtex4}
\usepackage{epsfig}
\usepackage{times}
\usepackage{color}

\begin{document}

\title{The evolution of lying in well-mixed populations}

\author{Valerio Capraro}
\email{v.capraro@mdx.ac.uk}
\affiliation{Department of Economics, Middlesex University, The Burroughs, London NW4 4BT, U.K.}

\author{Matja{\v z} Perc}
\affiliation{Faculty of Natural Sciences and Mathematics, University of Maribor, Koro{\v s}ka cesta 160, SI-2000 Maribor, Slovenia}
\affiliation{Complexity Science Hub Vienna, Josefst{\"a}dterstra{\ss}e 39, A-1080 Vienna, Austria}

\author{Daniele Vilone}
\affiliation{LABSS (Laboratory of Agent Based Social Simulation), Institute of Cognitive Science and Technology, National Research Council (CNR), Via Palestro 32, 00185 Rome, Italy}
\affiliation{Grupo Interdisciplinar de Sistemas Complejos (GISC), Departamento de Matem\'aticas, Universidad Carlos III de Madrid, 28911 Legan\'es, Spain}

\begin{abstract}
Lies can have profoundly negative consequences for individuals, groups, and even for societies. Understanding how lying evolves and when it proliferates is therefore of significant importance for our personal and societal well-being. To that effect, we here study the sender-receiver game in well-mixed populations with methods of statistical physics. We use the Monte Carlo method to determine the stationary frequencies of liars and believers for four different lie types. We consider altruistic white lies that favor the receiver at a cost to the sender, black lies that favor the sender at a cost to the receiver, spiteful lies that harm both the sender and the receiver, and Pareto white lies that favor both the sender and the receiver. We find that spiteful lies give rise to trivial behavior, where senders quickly learn that their best strategy is to send a truthful message, whilst receivers likewise quickly learn that their best strategy is to believe the sender's message. For altruistic white lies and black lies, we find that most senders lie while most receivers do not believe the sender's message, but the exact frequencies of liars and non-believers depend significantly on the payoffs, and they also evolve non-monotonically before reaching the stationary state. Lastly, for Pareto white lies we observe the most complex dynamics, with the possibility of both lying and believing evolving with all frequencies between 0 and 1 in dependence on the payoffs. We discuss the implications of these results for moral behavior in human experiments.
\end{abstract}

\maketitle

\section*{Introduction}

There are arguments and data in favor of the statement that we live safer, richer, and healthier than ever before~\cite{pinker2011better,pinker2019enlightenment}. But the gap between rich and poor is currently growing out of all reasonable proportions. And it is difficult to look away from the armed conflicts, hunger, and poverty without thinking that we ought to be able to do better. While we try our best to be compassionate, civilized, and social, and while there is an abundance of technological breakthroughs and innovations that make our lives better, many human societies are still seriously failing to meet the most basic needs of millions around the world~\cite{humans_14}. We are also dangerously depleting natural resources, our industries and ways of life are changing the climate, and we have fallen victim to echo chambers and misinformation, to the point that it is often impossible to discern truth from lies~\cite{garrett2009echo, del2016spreading}.  

Although the above-outlined issues are diverse and multifaceted, they do share one common property. Their solutions require cooperation. And we do cooperate -- in fact, we are champions of cooperation, to the point that we exercise ``SuperCooperation''~\cite{nowak_11}. But since natural selection in all of biology favors the fittest and the most successful individuals, there is still an innate selfishness in us that greatly challenges our cooperative drive. Cooperation is costly, and exercising it weighs down on individual well-being and prosperity. We therefore often succumb to the Darwin within, and we forget about less privileged others, and about future generations, and the health of our climate, and about many related issues that would require large-scale cooperation to be improved. Not surprisingly, understanding and promoting cooperation in human societies has once been declared one of the grandest challenges of the 21st century~\cite{kennedy_s05}, and scholars from disciplines as diverse as economics, psychology, sociology, biology, and anthropology have explored what factors favor people's cooperative behavior~\cite{trivers1971evolution, axelrod_84, ostrom2000collective, henrich_aer01, milinski2002reputation, fehr2002altruistic, gintis2003explaining, tomasello2005understanding, nowak_s06, bowles_11,rand2013human, capraro2013model}.

Methods of physics, in particular the Monte Carlo method and related approaches in statistical physics and network science~\cite{stanley_71, binder_88, estrada2012structure, boccaletti_pr14, kivela_jcn14, barabasi_16}, have also emerged as being very useful for studying many social phenomena. Statistical physics of social dynamics~\cite{castellano_rmp09}, of evolutionary games in structured populations~\cite{szabo_pr07, perc_bs10, perc_jrsi13, wang_z_epjb15}, of crime~\cite{orsogna_plr15}, of gossip~\cite{giardini2016evolution}, and of epidemic processes and vaccination~\cite{pastor_rmp15, wang_z_pr16}, are all examples of this exciting development, with human cooperation being no exception~\cite{perc_pla16, perc2017statistical}. However, empirical work has shown that cooperation is only one kind of a more general class of behaviors -- moral behaviors~\cite{capraro2018right}. This suggests that the same methods could be applied effectively to study the evolution of other types of moral behaviors as well~\cite{capraro2018grand}.

Using this as motivation, here we use methods of statistical physics to study the evolution of lying, or deception. Why deception? Deception has significant negative impacts on government, companies, and the society as a whole. For example, tax evasion costs approximately 100 billion a year to the US government alone~\cite{gravelle2010tax}, whereas, still in the USA, insurance fraud costs about 40 billion a year to insurance companies~\cite{insurance}. More recently, research has also focused on the spreading of fake news and misinformation~\cite{del2016spreading}, which, by favouring the emergence of inaccurate beliefs about the real state of the society, may represent a serious threat to democracy~\cite{pennycook2018prior}. Thus not surprisingly, studying dishonesty has a long history of interest among social scientists~\cite{mazar2008dishonesty, ariely2012honest, gino2009contagion, gino2011unable, shalvi2011justified, shalvi2012honesty, shalvi2015self, verschuere2011ease, biziou2015does, capraro2017does, capraro2018gender, erat2012white, gneezy2018lying, capraro2019time, fischbacher2013lies}, with the sender-receiver game being a popular theoretical paradigm to measure (dis)honesty~\cite{gneezy2005deception}.

In what follows, we re-introduce the sender-receiver game in a way that is appropriate to use with the Monte Carlo method, and we determine the stationary frequencies of liars and believers for four different lie types in well-mixed populations. In particular, we consider altruistic white lies, black lies, spiteful lies, and Pareto white lies, and we study in detail the dynamics that emerges as a result. As we will show, with spiteful lies in play senders and receivers both quickly learn that their best strategy is to send a truthful message and believe it, respectively. But for other types of lying, the dynamics becomes more nuanced. For example, for altruistic white lies and black lies, we will show that most senders lie while most receivers do not believe the sender's message, while for Pareto white lies, we will show that both lying and believing can evolve with any frequencies between 0 and 1. Our research thus adds a theoretically rigorous quantitative component to studying dishonesty, which has important implications for better understanding moral behavior in general, as well as provides pointers for devising innovative human experiments to test the theory.

\section*{The sender-receiver game}

Behavioral scientists have invented several tasks to measure people's (dis)honesty. The more popular ones are the die-rolling-paradigm~\cite{fischbacher2013lies}, the matrix task~\cite{mazar2008dishonesty}, the Philip Sidney game~\cite{smith1991honest}, and the sender-receiver game~\cite{gneezy2005deception}. In this work, we focus on the sender-receiver game, which is particularly suitable for the application of the Monte Carlo method, being a game with two players and (practically) two strategies, whereas the die-rolling-paradigm and the matrix task are both decision problems, with no strategic component, in which one person has to decide whether to lie for their benefit, or not. Moreover, the sender-receiver game allows us to study four different types of lies (black lies, spiteful lies, altruistic white lies, and Pareto white lies), whereas the Philip Sydney game, although strategically similar to the sender-receiver game, permits to study only black lies. In particular, we focus on the variant of the sender-receiver game introduced by Erat and Gneezy in~\cite{erat2012white}.

The game is as follows. There are two potential allocations of money between the sender and the receiver, Option A and Option B. The sender rolls a six-face dice and is the only one who sees the outcome. After looking at the outcome, the sender chooses a message to send to the receiver among six possible messages: ``The outcome was $i$'', with $i\in\{1,2,3,4,5,6\}$. After receiving the message, the receiver has to guess the true outcome of the dice roll. If the receiver guesses the true outcome, then Option A is implemented as a payment; if the receiver fails to guess the true outcome, then Option B is implemented. 

Although, in principle, this game has six strategies for each player, it can be reduced to a game with two strategies for each player in an obvious way. The sender has indeed essentially two strategies: he either tells the truth to the receiver about the outcome of the dice, or he lies. Similarly, also the receiver has essentially two strategies: she either believes the message sent by the sender, or not: if the receiver believes the sender, she reports the same number as the one sent by the sender; otherwise, if the receiver does not believe the sender, she draws randomly a number from the remaining five numbers of the dice.

Therefore, we can write the payoff matrix of the sender-receiver game as follows. Let $A=(a_R,a_S)$ and $B=(b_R,b_S)$ be the payoffs associated to Option A and Option B, respectively, where $S$ stands for the sender and $R$ stands for the receiver. If the number chosen by the receiver is equal to the actual outcome of the dice, the sender gets the payoff $a_S$, and the receiver gets the payoff $a_R$. Conversely, if the number chosen by the receiver is not equal to the actual outcome of the dice, the sender gets the payoff $b_S$, and the receiver gets the payoff $b_R$.

Without loss of generality, we can reduce the number of parameters from four to two by setting $a_S=a_R=0$. Finally, by setting $s=b_S$ and $r=b_R$, we can rewrite the game in a $2\times2$ matrix form, as follows
\begin{center}
\begin{tabular}{ |c|c|c| }
 \hline
  & B & N \\
 \hline
 T & $0,0$ & $s,r$ \\
 L & $s,r$ & $\frac{4}{5}s$ ,$\frac{4}{5}r$  \\
 \hline
\end{tabular}
\end{center}
where $T$ stands for ``Truth'', $L$ stands for ``Lie'', $B$ stands for ``Believe'', and $N$ stands for ``Not Believing''. The ratios $\frac{4}{5}$ come from the fact that, when the sender lies and the receiver does not believe the message sent by the sender, then the receiver does not guess the true outcome of the dice with probability $\frac{4}{5}$.

Following the taxonomy introduced by Erat and Gneezy~\cite{erat2012white}, we distinguish four types of lies, depending on the consequences in payoffs:
\begin{itemize}
    \item Pareto white lies are those that benefit both the sender and the receiver: $r,s>0$.
    \item Altruistic white lies are those that benefit the receiver at a cost to the sender: $r>0$, $s<0$.
    \item Black lies are those that benefit the sender at a cost to the receiver: $r<0$, $s>0$.
    \item Spiteful lies are those that harm both the sender and the receiver: $r,s < 0$.
\end{itemize}

We conclude this section by reporting the equilibrium analysis. If $r,s < 0$, there are two equilibria in pure strategies, $(T,B)$ and $(L,N)$, and one equilibrium in mixed strategies $(T/6 + 5L/6, B/6 + 5N/6)$ -- that is, the sender plays $T$ with probability $1/6$ and plays $L$ with probability $5/6$; analogous for the receiver. If $sr<0$ (i.e., if $r>0$ and $s<0$ or $s>0$ and $r<0$), then there are no equilibria in pure strategies and there is one equilibrium in mixed strategies, that is, again, $(T/6 + 5L/6, B/6 + 5N/6)$. Finally, if $s,r > 0$, there are two equilibria in pure strategies, $(T,N)$, $(L,B)$, and one equilibrium in mixed strategies, again, $(T/6 + 5L/6, B/6 + 5N/6)$. The cases $r=0$ and/or $s=0$ are trivial, because the corresponding player/s is/are indifferent between the strategies.

\begin{figure*}
\centerline{\epsfig{file=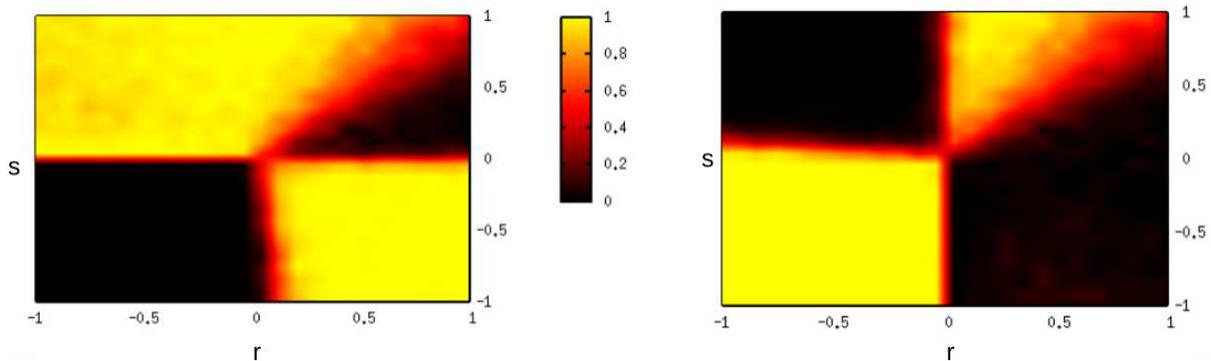,width=16.1cm}}
\caption{Density of liars (left panel) and believers (right panel) in the steady state. In the domain of spiteful lies, all senders are honest and all receivers believe the sender's message. In the domain of altruistic white lies and black lies, most senders lie and most senders do not believe the sender's message. However, the exact final frequencies depend on the specific parameters. In the domain of Pareto white lies, the steady state depends significantly on the parameter values. System size used is $N=500$, and the results are averaged over 2000 independent realizations.}
\label{liars_believers}
\end{figure*}

\section*{The Monte Carlo method}

We consider the sender-receiver game among $N$ players, who interact pairwise in a well-mixed population. At each round of the game, one player acts as a sender, and the other player acts as a receiver. Each player can assume either role, which is decided by a coin toss at the start of each encounter. When acting as a sender, a player can either tell the truth ($T$) or lie ($L$). When acting as a receiver, on the other hand, a player can either believe ($B$) the message received from the sender, or not ($N$). This gives rise to four different strategies, namely $(T,B)$, $(T,N)$, $(L,B)$, and $(L,N)$. Initially, each player is randomly assigned as either $T$ or $L$ (when she acts as a sender), and as either $B$ or $N$ (when she acts as a receiver).

We simulate the game using the Monte Carlo method. For a well-mixed population with $N$ players, the following elementary steps apply. First, a player $x$ is randomly drawn from the population. Player $x$ then plays the sender-receiver game with four randomly chosen other players from the population in a pairwise manner as described above, thereby obtaining the payoff $\pi_x$. Secondly, another player $y$ is also randomly drawn from the population, and he also plays the sender-receiver game with four randomly chosen other players from the population, thereby obtaining the payoff $\pi_y$. Lastly, player $y$ imitates the strategy of player $x$ in accordance with the probability $w=\{1+\exp[(\pi_{y}-\pi_{x})/K]\}^{-1}$, where $K$ quantifies the uncertainty during the strategy adoption process. In the $K \to \infty$ limit, payoffs cease to matter and strategies change at random; conversely, in the $K \to 0$ limit, player $y$ imitates $x$ only if $\pi_{x} > \pi_{y}$; between these two limits, the strategies of better performing players tend to be imitated, although under-performing strategies are imitated as well, for example due to errors in the decision making, imperfect information, and external influences that may adversely affect the evaluation of the payoff of the other player. Without loss of generality, here we  set $K=0.1$, in agreement with previous research that showed this to be a representative value~\cite{perc2017statistical}.

The time is measured in Monte Carlo steps (MCS), whereby one MCS corresponds to executing all three elementary steps $N$ times. During one MCS, each player changes strategy, on average, only once. For a systematic numerical analysis, we have determined the fraction of strategies in the final stationary state when varying the values of $s$ and $r$. For an adequate accuracy, we have used sufficiently large system sizes, varied from $N=500$ to $1000$, as well as long enough thermalization and sampling times, varied from $10^4$ to $10^6$ MCS. To further remove statistical fluctuations, we have also averaged the final outcome over up to $2000$ independent realizations.

\section*{Results}

We considered a well-mixed population and investigated the final configuration reached by the system once the dynamics has reached its steady state. 

\subsection*{Final densities of liars and believers as a function of lie type}

As a first step of our analysis, we look at the final densities of liars and believers, as functions of the type of lie.

\begin{figure*}
\centerline{\epsfig{file=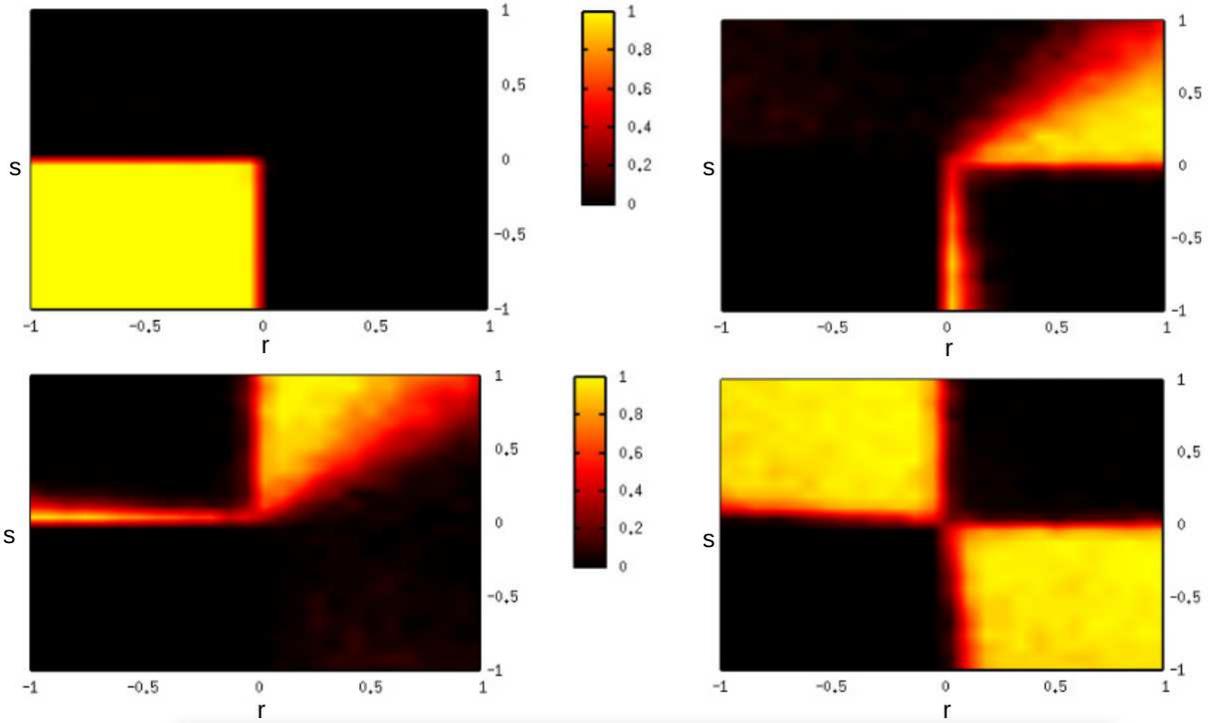,width=16.1cm}}
\caption{Upper-left panel: Final densities of the pure strategy profile $(T,B)$, which turns out to evolve only in the domain of spiteful lies. Upper-right panel: Final densities of the pure strategy profile $(T,N)$, which turns out to evolve in three cases, namely, for altruistic white lies, for black lies, and for Pareto white lies, although with different frequencies depending on the exact parameter values. Lower-left panel: Final densities of the pure strategy profile $(L,B)$, which also turns out to evolve in the domains of altruistic white lies, black lies, and Pareto white lies, but with different frequencies depending on the exact parameter values. Lower-right panel: Final densities of the pure strategy profile $(L,N)$, which turns out to evolve only in the domains of altruistic white lies and black lies, and, in both cases, with very high frequencies.} 
\label{density_pure}
\end{figure*}

Figure~\ref{liars_believers} shows the final densities of liars (left panel) and believers (right panel), as functions of the game parameters $(r,s)$. For each couple $(r,s)$, the corresponding densities are obtained by averaging over 2,000 independent realizations on a system of size $N=500$. The simulations were conducted with $r,s$ increasing from $-1$ to $1$, with steps of length $0.08$. We verified that the dynamics has actually reached the final state. 

We start from the case $r,s<0$. The left panel highlights that, in this case, all senders are honest, whereas the right panel puts in evidence that all receivers believe the message sent by the sender. This result is not a priori obvious. The case $r,s<0$ corresponds to spiteful lies, in which both the sender and the receiver are harmed by a lie that is believed. As we have seen before, in this domain, the sender-receiver game has three equilibria $(T,B)$, $(L,N)$, and $(\frac16T+\frac56L,\frac16B+\frac56N)$. The simulations show that two of these equilibria are discarded and all agents tend to coordinate on $(T,B)$. A theoretical reason for why this happens is that this equilibrium is the only one that is Pareto optimal in that it maximizes the payoff for both players. Therefore, $(T,B)$ is the strategy that has the most chances to be imitated. Also note that, as shown in this figure (see also the upper-left panel of Figure 2), the finding that only the $(T,B)$ equilibrium survives in the evolution is robust to changing the payoff parameters, $r$ and $s$, as long as they remain in the domain of spiteful lies. In other words, in the domain of spiteful lies, senders quickly learn that their best strategy is to report the truth, while receivers quickly learn that their best strategy is to believe the sender's message. 

Now, keeping $r<0$ constant, we note that, when $s$ increases and overcomes zero, there is a state transition, which corresponds to the fact that the parameters $(r,s)$ enter the domain of black lies, where, assuming that receivers believe the senders' messages, it is favorable for senders to lie. This has the effect that lying tends to spread. However, since, in the domain of black lies, receiver's best response to lying (L) is to not believe the sender's message (N), while L emerges, also N emerges. The emergence of N in turn contrasts the emergence of L among senders, because, in the domain of black lies, senders' best response to N is telling the truth (T). This opposite dynamics result in a mixed steady state in which most, but not all, senders lie, and most, but not all, receivers, do not believe the sender's message. One might at this point wonder whether this stationary state is equal to the unique mixed strategies equilibrium, and, in particular, whether it is independent of the parameters $(r,s)$, or not. The answers are negative. We will show in the next sections that, in fact, the steady state depends on the parameters $(r,s)$ non-trivially.

A similar logic applies when we keep $s<0$ and let $r$ increase from $-1$ to  $1$. As soon as $r$ becomes positive, there is a state transition corresponding to the fact that the parameters $(r,s)$ enter the domain of altruistic white lies. In this domain, assuming that receivers believe that senders tell the truth, then it is favorable for receivers to not believe the sender's message. This has the effect that strategy N tends to emerge. However, since in the domain of altruistic white lies, sender's best response to N is L, the emergence of N is contrasted by the emergence of L. This opposite dynamics result in a mixed state state, which, again, depends non-trivially from the exact parameters $(r,s)$ as we will show in the next sections.

The quadrant in which both $r$ and $s$ are positive is the more variegate one. These parameters correspond to Pareto white lies, lies that benefit both the sender and the receiver. The resulting dynamics is quite complex and the steady state highly depend on the parameters $(r,s)$, and both L and N can span all possible frequency values from 0 to 1, in a monotonic way: keeping $r$ constant, the final frequencies of L and B both increase with $s$.

\begin{figure*}
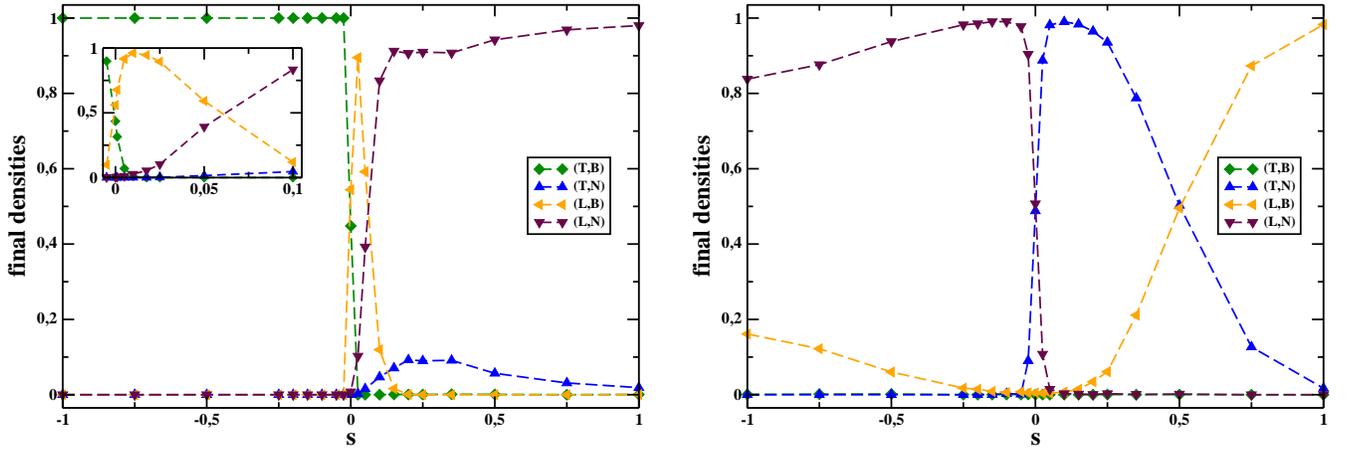

\centerline{\epsfig{file=figure3a.eps, width=8.5cm}\hspace{0.6cm}\epsfig{file=figure3b.eps, width=8.5cm}}
\caption{Left panel: Final densities of different strategies as a function of the parameter $s$, for fixed $r=-0.50$. When $s<0$, only the strategy $(T,B)$ survives. For $s>0$, $(L,B)$ quickly increases to around 0.9 and then it quickly decreases to 0; $(L,N)$ quickly increases up to around 0.9, and then slowly keeps increasing up to reaching values close to 1; $(T,B)$ completely vanishes; $(T,N)$ first emerges for small values of $s$, then vanishes; inset: zoom of the interval $s\in[-0.005,0.1]$. Right panel: Final densities of the different strategies as a function of the parameter $s$, for fixed $r=0.50$. For $s<0$, only $(L,B)$ and $(L,N)$ emerge, although the latter with much higher probability. For $s>0$, $(T,N)$ quickly emerges, but then it slowly disappears, contrasted by the emergence of $(L,B)$. In all cases the system size used is $N=1000$ with random initial conditions.}
\label{believ1}
\end{figure*}

\begin{figure*}
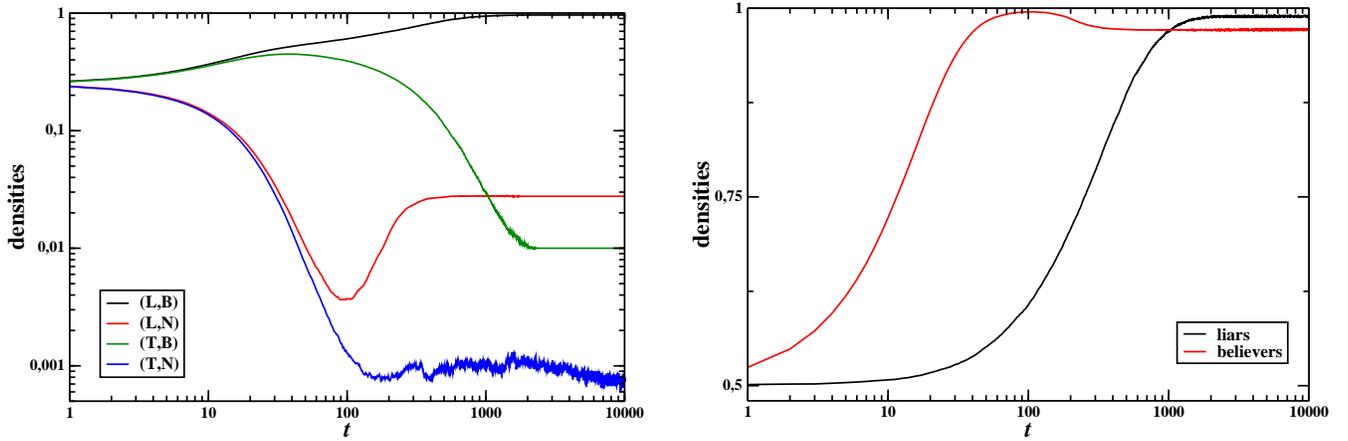

\centerline{\epsfig{file=figure4a.eps, width=8.5cm}\hspace{0.6cm}\epsfig{file=figure4b.eps, width=8.5cm}}
\caption{Left panel: Time series of the frequencies four basic strategies for $r=-1$ and $s=0.01$, that is, around the $(L,B)$ maximum, see Figure~\ref{believ1}~(left). The frequency of $(L,B)$ increases monotonically up to near 1, while all other strategies tend to appear with very small frequencies, although their evolution is rather different. In particular, $(L,N)$ evolves non-monotonically, while $(T,N)$ is even oscillatory.  Right panel: Time series of the frequencies of liars and believers for the same parameter values utilized in the left panel. System of size $N=1000$ with random initial conditions.}
\label{zoom_dyn}
\end{figure*}

\subsection*{Density of the pure strategies}

In the previous section, we have reported the final densities of liars and believers as a function of the type of lie. However, liars can come in two forms: liars who, when playing in the role of the receiver, believe the sender's message and liars who, when playing in the role of the receiver, do not believe the sender's message. Similarly, believers can come in two forms: believers who, when playing in the role of the sender, send a truthful message and believers who, when playing in the role of the sender, send a deceptive message. To gain insights about which strategies are more likely to evolve, in this section we report and discuss the final densities of the four pure strategy profiles $(T,B)$, $(T,N)$, $(L,B)$ and $(L,N)$.

The upper-left panel of Figure~\ref{density_pure} highlights that the strategy profile $(T,B)$, according to which a player reports the truth when acting as a sender and believes the sender's message when acting as a receiver, appears in the steady state only for $r,s<0$ (spiteful lies). In all other types of lie, the pure strategy profile $(T,B)$ never evolves.

Particularly interesting is the strategy profile $(T,N)$, according to which a player tells the truth when acting as a sender, but does not believe the sender's message, when acting as a receiver. This situation is similar to what Sutter~\cite{sutter2009deception} termed ``sophisticated deception'', telling the truth while expecting to not be believed. The upper-right panel of Figure~\ref{density_pure} highlights that this strategy profile appears in a number of non-trivial cases. When $s$ is negative and $r$ is positive and close to zero $(T,N)$ appears with high probability, close to 1. This case corresponds to altruistic white lies that have a very small cost for the sender. Instead, when $r$ is negative and $s$ is positive (black lies), $(T,N)$ emerges, but it does so with very small probability. In the domain of Pareto white lies ($r,s>0$), $(T,N)$ almost always emerges (especially for $r\geq s$). In particular, when $r$ gets close to 1 and $s$ is between $0$ and $0.5$, $(T,N)$ emerges with very high probability, close to 1.

The case $(L,B)$ is symmetric to the case $(T,N)$. The lower-left panel of Figure~\ref{density_pure} shows that this strategy profile does not emerge at all in the domain of spiteful lies ($r<0, s<0$) and it emerges with small probability in the domain of altruistic white lies ($r>0, s<0$). In the domain of black lies ($r <0, s>0$), we note a fast emergence of the strategy $(L,B)$ for small values of $s$, close to $0$, in which this strategy profile evolves even with probability close to 1. However, for larger values of $s$ it quickly vanishes. Again, the domain of Pareto white lies is the more variegate one. Indeed, in this case, the strategy profile $(L,B)$ emerges with high probability when $s\geq r$, whereas for $s<r$, its probability is very small.

Finally, the lower-right panel of Figure~\ref{density_pure} shows that the strategy profile $(L,N)$ does not emerge in the domains of spiteful lies and Pareto white lies, but it does emerge in the domains of altruistic white lies and black lies, with very high, although not equal to 1, probabilities.

\begin{figure*}
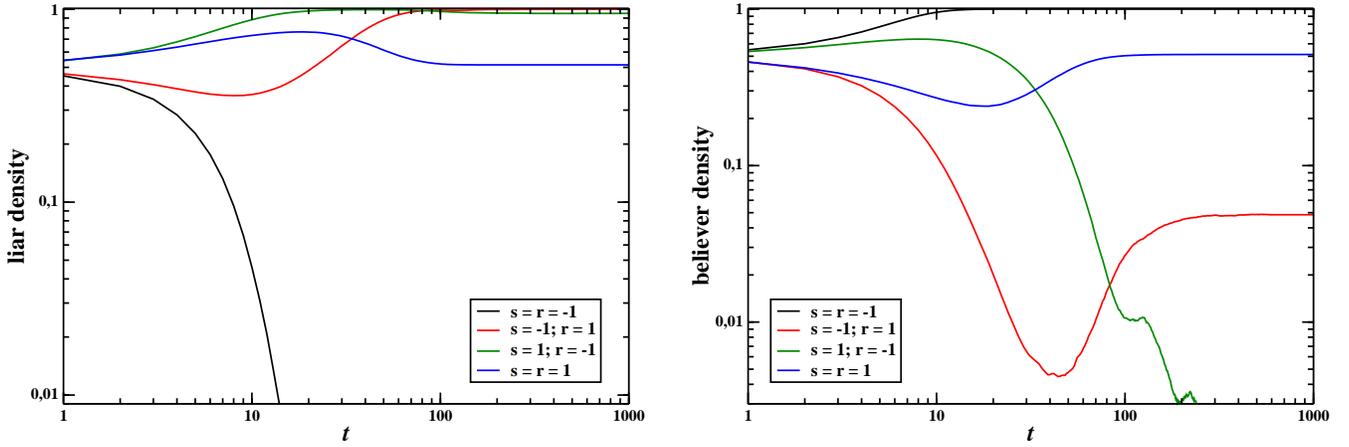

\centerline{\epsfig{file=figure5a.eps, width=8.5cm}\hspace{0.6cm}\epsfig{file=figure5b.eps, width=8.5cm}}
\caption{Left panel: Time evolution of liars at the corners of the domain of the parameters $(r,s)$. The evolution is monotone only in the case of $r=-1$ and $s=-1$ (spiteful lies). Right panel: Time evolution of believers at the corners of the domain of the parameters $(r,s)$. Time evolution of liars at the corners of the domain of the parameters $(r,s)$. The evolution is monotone only in the case of $r=-1$ and $s=-1$ (spiteful lies). In all cases the system size used is $N=500$ with random initial conditions.}
\label{densities_liars_believers}
\end{figure*}

\begin{figure*}
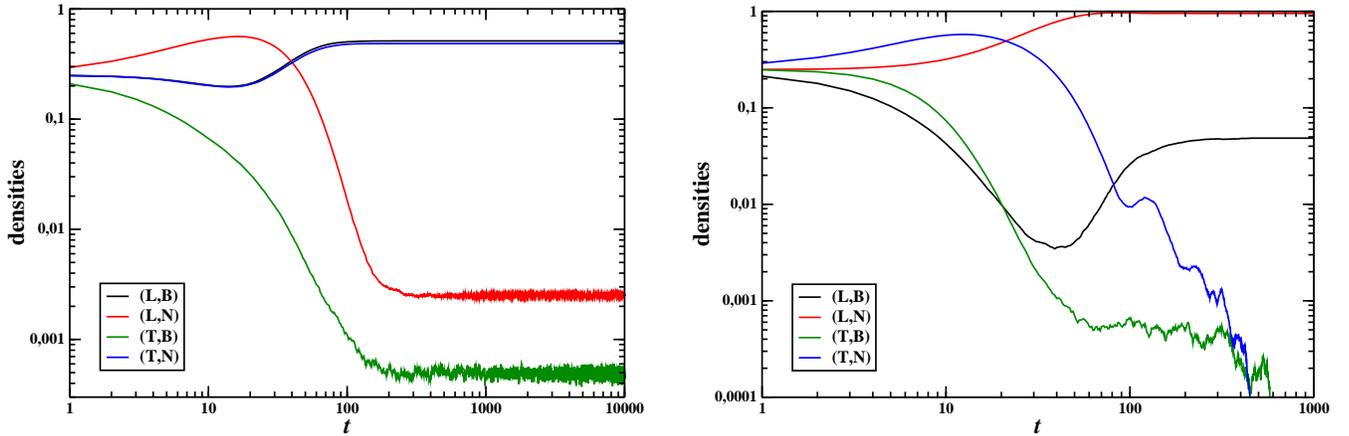

\centerline{\epsfig{file=figure6a.eps, width=8.5cm}\hspace{0.6cm}\epsfig{file=figure6b.eps, width=8.5cm}}
\caption{Left panel: Time evolution of the four pure strategy profiles for $r=1$ and $s=1$ (Pareto white lies). The non-monotonic evolution of liars is primarily driven by a non-monotonic evolution of the strategy $(L,N)$. The non-monotonic evolution of believers is driven by a combination of $(T,B)$ and $(L,B)$. Right panel: Time evolution of the four pure strategy profiles for $r=1$ and $s=-1$ (altruistic white lies). The non-monotonic evolution of liars and believers is mainly driven by a non-monotonic evolution of the strategy $(L,B)$. Systems of size $N=500$ with random initial conditions.}
\label{densities_strategies}
\end{figure*}

\subsection*{Sections}

We have said earlier that, in the domains of black lies ($r<0$, $s>0$) and altruistic white lies ($r>0$, $s<0$), the steady state depends on the specific values of $r$ and $s$ in a non-trivial way, and that, in particular, it is not equal to the unique Nash equilibrium of the game, $(\frac16T+\frac56L,\frac16B+\frac56N)$. Here we show this interesting fact by reporting the dynamics along the two sections $r=\pm0.50$, as functions of the sole parameter $s$.

We start by setting $r=-0.50$. When $s<0$, we have already seen in the previous section that the only strategy profile that survives is $(T,B)$. This is indeed reflected in Figure~\ref{believ1} (left), which puts in evidence that, in this region, the frequency of $(T,B)$ (green line) is equal to 1, whereas all other frequencies are equal to 0. Then, when $s$ becomes positive, there is a sudden change of state. Interestingly, liars quickly emerge, but in a non-symmetric way: the frequency of $(L,B)$ quickly increases up to almost 1 for $s\simeq0.01$, as shown in the inset of Figure~\ref{believ1} (left), then it quickly decreases again to 0. On the other hand, the frequency of $(L,N)$ rapidly increases up to around 0.9, and then slowly keeps increasing up to reaching a value near 1. The maximum of the frequency of $(L,B)$ is rather surprising for its narrowness: the final density of $(L,B)$ is 0 for $s<0$; then it quickly increases for positive but very small values of $s$; then it quickly decreases again to 0. To better understand this peculiar behavior, in Figure~\ref{zoom_dyn} we report the time series of each strategy in the interval of $(L,B)$ dominance. Specifically, the left panel of Figure~\ref{zoom_dyn} highlights that the frequency of $(L,B)$ increases monotonically up to near 1, while all other strategies tend to appear with very small frequencies, although their evolution is rather different. In particular, $(L,N)$ evolves non-monotonically, while (T,N) is even oscillatory. The right panel of Figure~\ref{zoom_dyn} reports the evolution of liars and believers in the same interval of $(L,B)$ dominance. (More details about the time evolution of the various densities will be given in the next section.)
Regarding truth-telling, the strategy $(T,B)$, which was the only surviving strategy for $s<0$, in the domain $s>0$ completely vanishes. On the other hand, the strategy $(T,N)$ emerges in a non-monotonic way: as $s>0$ increases, the frequency of $(T,N)$ first increases up to a value around 0.1, and then slowly decreases to values near 0. Therefore, for $r=-0.5$ and $s>0$, receivers never believe the sender's message, while senders lie with high frequency, but not equal to 1.

The case $r=0.50$ is somewhat more articulated, as shown in Figure~\ref{believ1} (right). When $s<0$, liars emerge with frequency 1, however, this does not appear to be due to the emergence of a single profile of strategies. Indeed, for $s<0$ we see a coexistence of the strategy profiles $(L,B)$ and $(L,N)$, although the latter one appears to emerge with higher frequency, especially when $s$ increases and approaches $0$, in which $(L,N)$ reaches frequencies very close to 1. Then, as soon as $s$ reaches $0$, there is a change of state: the strategy profile $(T,N)$ appears with frequency very close to 1; however, as $s$ increases towards 1, then $(T,N)$ appears with lower and lower frequencies. This decrease of the frequency of appearance of $(T,N)$, as $s$ increases, appears to be perfectly mirrored by an increase of the frequency of $(L,B)$.  

\subsection*{Time evolution}

We conclude by reporting the time evolution of liars and believers at the corner of the domain of the parameters $(r,s)$. We verified the time evolution also for other values of $(r,s)$, and we found qualitatively similar patterns (as long as $r,s\neq0$, clearly). 

Figure~\ref{densities_liars_believers}, left and right, highlight that, before reaching the steady state, the evolution is interesting, being sometimes monotone and sometimes not. For $r=1$ and $s=-1$ (red line, altruistic white lie), we note that both the behavior of senders and the behavior of receivers evolve in a non-monotone way. Similarly, for $r=1$ and $s=1$ (blue line, Pareto white lie), the behavior of both senders and receivers evolve non-monotonically. A non-monotone evolution, although less remarked, appears also in the case of black lies ($r=-1$, $s=1$, green line). Conversely, in the case of spiteful lies, we see a very quick convergence to the strategy $(T,B)$, in line with the discussion above that, in this case, senders quickly learn that their best strategy is to tell the truth and receivers quickly learn that their best strategy is to believe the sender's message.

Figure~\ref{densities_strategies} reports in more detail the time evolution of the four basic strategies for $r=1, s=\pm1$, that is, when the densities of liars and believers evolve non-monotonically. In the case of Pareto white lies (left panel), we note that the non-monotonic evolution of liars is primarily driven by a non-monotonic evolution of the strategy $(L,N)$, whose frequency first increases up to about $0.8$ and then suddenly decreases of two orders of magnitudes, to values below $0.01$, and then keeps oscillating. Similarly, still in the domain of Pareto white lies, the non-monotonic evolution of believers is driven by a combination of $(T,B)$ and $(L,B)$: at the beginning of the dynamics, the frequency of $(L,B)$ is approximately constant, while the frequency of $(T,B)$ decreases, giving rise to the initial decrease of believers observed in the right panel of Figure~\ref{densities_liars_believers}; then, between $t\simeq20$ and $t\simeq100$, the frequency of $(T,B)$ doubles from about $0.4$ to about $0.8$, where it stabilizes, while the frequency of $(T,B)$ keeps decreasing. After $t\simeq100$, the frequency of $(T,B)$ starts alternating. This change in the dynamics contributes to the overall non-monotonicity observed in the evolution of the frequency of believers. A similar line of reasoning holds in the case of altruistic white lies. As shown in the right panel of Figure~\ref{densities_strategies}, the non-monotonic evolution of liars and believers is mainly driven by a non-monotonic evolution of the strategy $(L,B)$.

Finally, it is worth noticing that the non-monotonic behaviour in time increase with the population size: indeed, for very large systems ($N\gtrsim10^4$), in some cases we observe oscillations before the densities reach the final state.

\subsection*{Discussion}

We have used the Monte Carlo method to explore the evolution of lying in well-mixed populations, where individuals are playing the sender-receiver game~\cite{gneezy2005deception, erat2012white}. We have shown that the evolution follows non-trivial trajectories. In particular, honesty and dishonesty may appear or disappear with very high probability depending on the particular payoffs of the game. Similarly, also believing and non-believing can emerge or vanish with very high probabilities. More specifically, following Erat and Gneezy's taxonomy of lies~\cite{erat2012white}, we distinguished four basic types of lies: black lies, spiteful lies, altruistic white lies, and Pareto white lies. In the domain of spiteful lies, senders quickly learn that their best strategy is to send a truthful message, and receivers quickly learn that their best strategy is to believe the sender's message. The cases of altruistic white lies and black lies are instead characterized by the fact that, at the steady state, most senders lie while most receivers do not believe the sender message. However, the exact proportions of senders and non-believers depend significantly on the particular payoffs, and they also evolve in a non-monotonic way, before eventually reaching the steady state. The case of Pareto white lies is an even more variegate one. Here, the steady state depend fully on the payoffs, and both lying and non-believing can evolve with all probabilities between 0 to 1.

Previous research has explored the evolution of honesty using the Philip Sidney game~\cite{smith1991honest}. In this game, the Sender is initially in either of two states, healthy or needy, with probability $p$ and $1-p$, respectively. The Sender can either pay a cost $c$ to signal his state or stay quiet. The Receiver does not know the state of the Sender, but can observe the signal. After observing the signal (if sent), the Receiver decides whether to donate his resource to the Sender. The Sender and the Receiver are assumed to be related, by a relatedness coefficient $r$. Each player's payoff is the sum of his survival probability and a fraction $r$ of the other player's survival probability. Survival probabilities are defined as follows: the Receiver is sure to survive only if he does not donate his resource; the Sender is sure to survive only if he receives the Receiver’s resource. This creates a conflict of interests among the Sender and the Receiver which corresponds to what we called (following Erat and Gneezy \cite{erat2012white}) the ``black lie'' condition. A classic work on the Philip Sidney game found that, if the cost of the signal is sufficiently high, then honest signalling can evolve~\cite{grafen1990biological}. See~\cite{szamado2011cost} for a review of this ``Handicap Principle'' and its variants. More recent research revealed that punishment can promote the evolution of honesty in cases in which the conflict of interests among the Sender and the Receiver is moderate and signalling is cheap or even cost-free~\cite{catteeuw2014evolution}. Our work departs from this line of research along two main dimensions. First, in the Sender-Receiver game, signalling is cost-free and there is no punishment. Even in this case, our results indicate that honesty can evolve in some circumstances (especially in the case of spiteful lies and Pareto white lies, but also, to some extent, in the case of black lies). Second, the Sender-Receiver game allows to study the evolution of honesty not only in the domain of black lies, but also in the domains of spiteful lies, Pareto white lies, and altruistic white lies.

Related to our work is also the recent literature on pre-commitments in social dilemmas. In this context, a social dilemma is preceded by a pre-play stage in which players can send messages (commitment proposals) and other players can accept or refuse the proposal. Proposers can lie about the commitment. For example, after promising that they would cooperate, proposers can dishonour their promise and defect. On the other hand, responders can refuse a commitment proposal because they do not believe the proposer. Han and colleagues explored analytically and numerically the evolution of cooperation in this type of social dilemmas, both in pairwise~\cite{han2013good} and group interactions~\cite{han2015avoiding,pereira2017evolution}, and found that cooperation can evolve under a number of different circumstances, such as for example when the cost of commitment is sufficiently small compared to the cost of cooperation. Our work differs from this line of research in that we focus specifically on honesty and believing, with no consequences on cooperative behaviour. This allows us to clearly identify the four classes of lies (black, spiteful, altruistic, Pareto), and to study the evolution of lying as a function of lie type.

Statistical physics, and, in particular, the Monte Carlo method, has proven valuable for the study of the evolution of cooperation in social dilemmas~\cite{perc2017statistical}. Yet, cooperation in social dilemmas is only one particular instance of a more general class of behaviors, moral behaviors~\cite{capraro2018right}. Therefore, it is time now to move beyond the borders of cooperation and start applying similar methods to the evolution of other moral behaviors, such as, indeed, honesty~\cite{capraro2018grand}. To the best of our knowledge, this is the first study using techniques from statistical physics to study the evolution of lying in the six-dice sender-receiver game. Of course, some questions remain to be addressed in future research, such as: What happens for general $n$-dice sender-receiver games? What happens on networks? What interventions can be done to favor the evolution of honesty? What if imitation is replaced with other forms of strategy change? Just to name a few. These are important questions, whose answers can greatly contribute to the improvement of the society we live in, and they can provide a nuanced quantitative view of honest behavior, as well as inform the design of future human experiments with testable theoretical predictions. 

Extending the domain of application of the Monte Carlo method from cooperation to honesty, our work also suggests that similar techniques could be applied to study the evolution of other forms of moral behavior. A recent work by Curry {\it et al}.~\cite{curry2017good} shows that seven moral rules are universal across societies: love your family, help your group, return favors, be brave, defer to authority, be fair, and respect others' property. Clearly, not all these behaviors can be studied using simple games, but some are. For instance, ``returning favors'' could be studied using a sequential prisoner's dilemma or the trust game; ``help your group'' could be studied using games with labeled players, in which individuals come with a label describing the group they belong to; ``fairness'' could be studied through the ultimatum game, as indeed has already been done~\cite{szolnoki2012defense,page2000spatial,kuperman2008effect,eguiluz2009critical,da2009statistical,deng2011coevolutionary,gao2011coevolutionary,szolnoki2012accuracy,deng2012network,iranzo2012empathy,miyaji2013evolution}; “respect others' property” can be studied utilizing games with special frames, as, for example, the dictator game in the take frame, for which taking turns out to be considered more morally wrong than giving~\cite{krupka2013identifying,capraro2019power}.

In sum, we believe that illuminating if, when, and how techniques of statistical physics can be applied to study the evolution of morality among humans, should be considered as a primary direction for future research.

\begin{acknowledgments}
Matja{\v z} Perc was supported by the Slovenian Research Agency (Grants J4-9302, J1-9112 and P1-0403). Daniele Vilone was supported by the European Union's Horizon 2020 Project PROTON (Grant 699824).
\end{acknowledgments}

\bibliography{refs}

\end{document}